\documentclass[
]{ceurart}

\begin{document}

\copyrightyear{2021}
\copyrightclause{Copyright for this paper by its authors.
  Use permitted under Creative Commons License Attribution 4.0
  International (CC BY 4.0).}


\conference{Proceedings of the Demonstration {\&} Resources Track, Best BPM Dissertation Award, and Doctoral Consortium at BPM 2021 co-located with the 19th International Conference on Business Process Management, BPM 2021, Rome, Italy, September 6-10, 2021}

\title{PC4PM: A Tool for Privacy/Confidentiality Preservation in Process Mining}

\author{Majid Rafiei}[%
orcid=0000-0001-7161-6927,
email=majid.rafiei@pads.rwth-aachen.de,
]
\author{Alexander Schnitzler}[%
orcid=0000-0002-8223-6097,
email=Alexander.Schnitzler@outlook.com,
]

\author{Wil M.P. van der Aalst}[%
orcid=0000-0002-0955-6940,
email=wvdaalst@pads.rwth-aachen.de,
]

\address{Chair of Process and Data Science, RWTH Aachen University, Aachen, Germany}

\begin{abstract}
Process mining enables business owners to discover and analyze their actual processes using event data that are widely available in information systems. 
Event data contain detailed information which is incredibly valuable for providing insights. However, such detailed data often include highly confidential and private information. Thus, concerns of privacy and confidentiality in process mining are becoming increasingly relevant and new techniques are being introduced. 
To make the techniques easily accessible, new tools need to be developed to integrate the introduced techniques and direct users to appropriate solutions based on their needs.
In this paper, we present a Python-based infrastructure implementing and integrating state-of-the-art privacy/confidentiality preservation techniques in process mining. 
Our tool provides an easy-to-use web-based user interface for privacy-preserving data publishing, risk analysis, and data utility analysis. 
The tool also provides a set of anonymization operations that can be utilized to support privacy/confidentiality preservation. The tool manages both standard XES event logs and non-standard event data. We also store and manage privacy metadata to track the changes made by privacy/confidentiality preservation techniques.
\end{abstract}


\maketitle

\section{Introduction}\label{sec:introduction}
Process mining techniques employ event logs to provide insights into actual processes \cite{van2016process}.
Event logs contain detailed information about operational processes and can be extracted from various types of information systems, e.g.,  EPR systems. 
Events are considered as the smallest units of process execution which are distinguished by their attributes. 
The main attributes of events are as follows: \textit{activity}, \textit{case id}, \textit{timestamp}, and \textit{resource}. 
For instance, a heart surgery (\textit{activity}) performed by Dr. John (\textit{resource}) for a patient with id=10 (\textit{case id}) at timestamp 2021.06.10-10:00:00 is an event recorded by an information system in a hospital. 
The attributes that directly or indirectly refer to individuals raise privacy concerns.
For example, in the healthcare context, the \textit{case id} attribute may refer to the patients whose data are processed, and the \textit{resource} attribute may refer to the employees who perform activities for the patients, e.g., nurses. 
Furthermore, other attributes can also be considered as confidential information, e.g., the \textit{activity} attribute may contain a confidential activity name that must not be exposed.
Respect for privacy when analyzing personal data is also dictated by regulations, e.g., the European General Data Protection Regulation (GDPR)\footnote{http://data.europa.eu/eli/reg/2016/679/oj}. 
Such legitimate and ethical requirements have recently resulted in more attention to privacy and confidentiality issues in process mining \cite{pripel_short,rafiei_group_arxiv,rafiei_challenge_arxiv,smcProcessMining_short}.
Some tools have also been introduced to provide specific privacy/confidentiality requirements \cite{shareprom_short,rafieippdpTool_arxiv,elpaas}.

\begin{wrapfigure}{r}{0.5\textwidth} 
	\centering
	\includegraphics[width=0.5\textwidth]{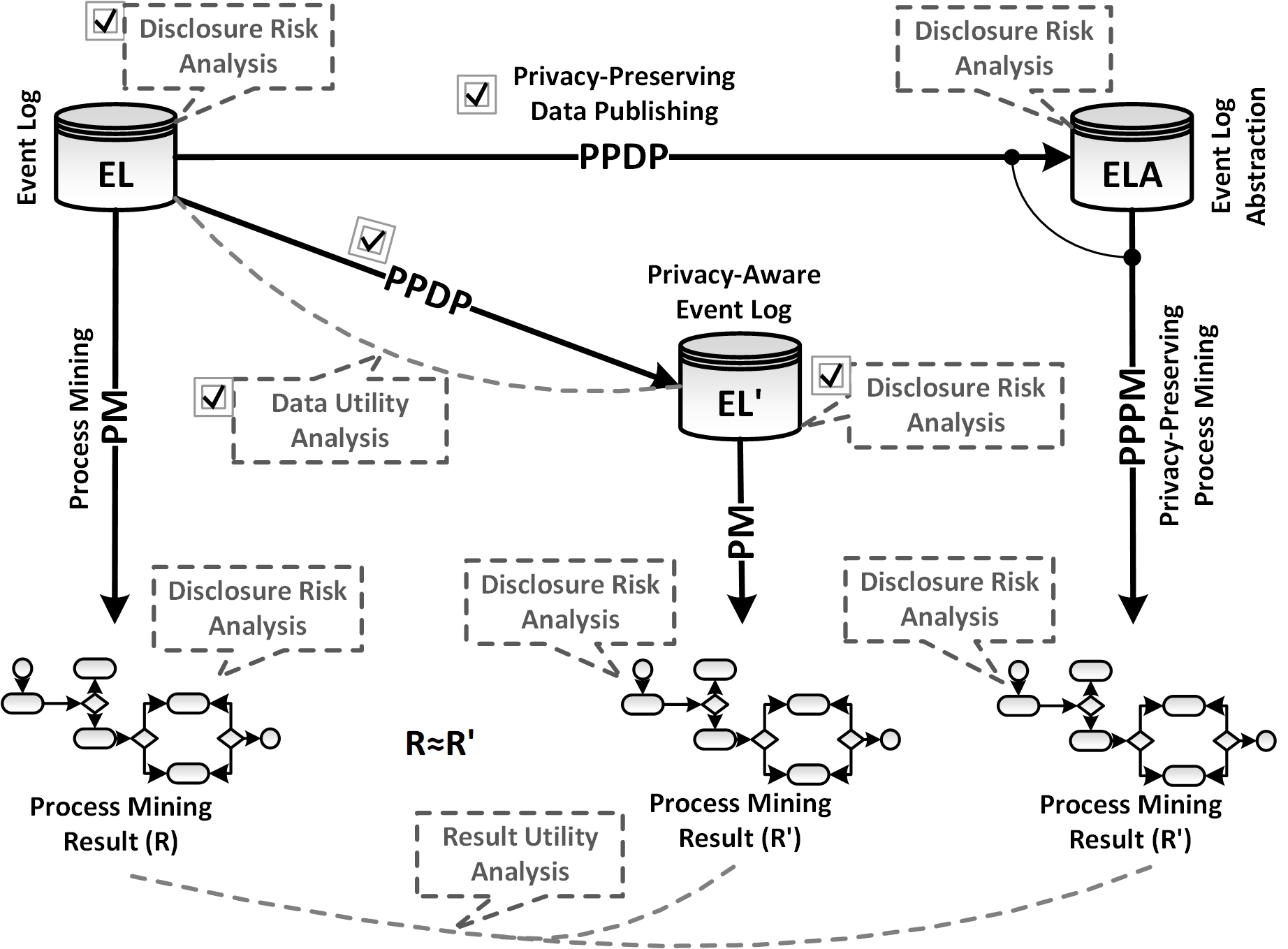}
	\caption{The general overview of privacy-related activities in process mining.}\label{fig:overview}
\end{wrapfigure}

Figure~\ref{fig:overview} shows the general overview of privacy-related activities in process mining including Privacy-Preserving Data Publishing (PPDP), Privacy-Preserving Process Mining (PPPM), and Privacy Analysis (PrAn). 
PPDP tries to obscure the identity and/or sensitive data of individuals to preserve their privacy. PPDP techniques often apply one or more \textit{anonymization operations}, e.g., \textit{suppression}, \textit{generalization}, etc., to provide the desired privacy requirements. 
PPPM intends to expand existing process mining algorithms to cope with intermediate results, so-called \textit{abstractions} \cite{rafieippdp_google}, generated by some PPDP techniques. 
Note that PPPM algorithms are closely linked with the corresponding PPDP approaches, and PPPM may refer to the entire privatization process, starting with an event log and finishing with process mining findings. 
PrAn, indicated with dashed lines in Figure~\ref{fig:overview}, includes two types of activities: \textit{risk analysis} and \textit{utility analysis}. Both PrAn activities could be done for data and results. In this paper, we introduce a tool, named PC4PM, mainly focusing on the activities indicated by the check-boxes in Figure~\ref{fig:overview}. PC4PM is the successor of the privacy tool introduced in \cite{rafieippdpTool_arxiv}, and it offers new privacy preservation techniques, privacy analysis, a set of anonymization operations, and user guidance that directs users to the right techniques based on their requirements. 
In the rest of the paper, we demonstrate the functionality and characteristics of PC4PM. We also describe the maturity and availability of the tool.

\section{Functionality and Characteristics}\label{sec:function}

\begin{wrapfigure}{r}{0.4\textwidth} 
	\centering
	\includegraphics[width=0.4\textwidth]{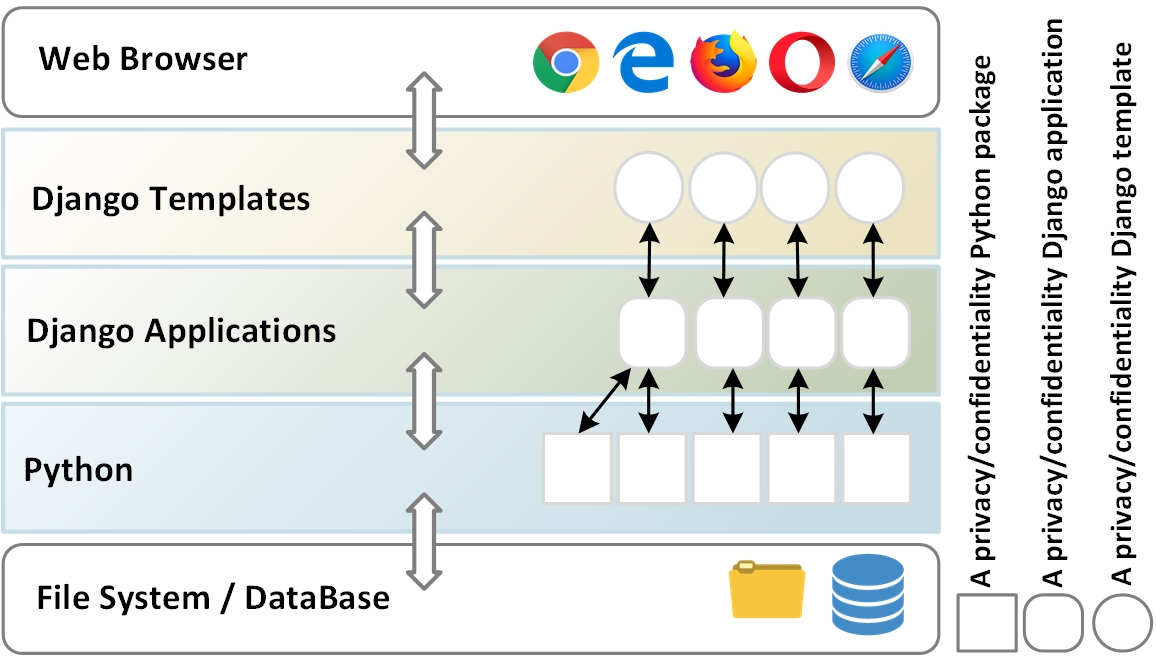}
	\caption{The architecture of PC4PM.}\label{fig:arch}
\end{wrapfigure}

PC4PM is implemented in Python using Django framework\footnote{https://www.djangoproject.com/}. Figure~\ref{fig:arch} shows a high-level view of the architecture. 
PC4PM includes eight main Django applications and each application provides at least one main privacy-related activity implemented as a Python package. The Django templates, accessible from any web browser, provide a web interface for the applications.
Implementing each technique as an independent Django application enables users to simultaneously run different techniques on event logs. Such architecture makes the process of maintenance and integration simple. To integrate new techniques, one can create a Python package and integrate it as an independent application. Moreover, Python packages can independently be imported into other Python-based tools. 

\begin{wrapfigure}{r}{0.55\textwidth} 
	\centering
	\includegraphics[width=0.55\textwidth]{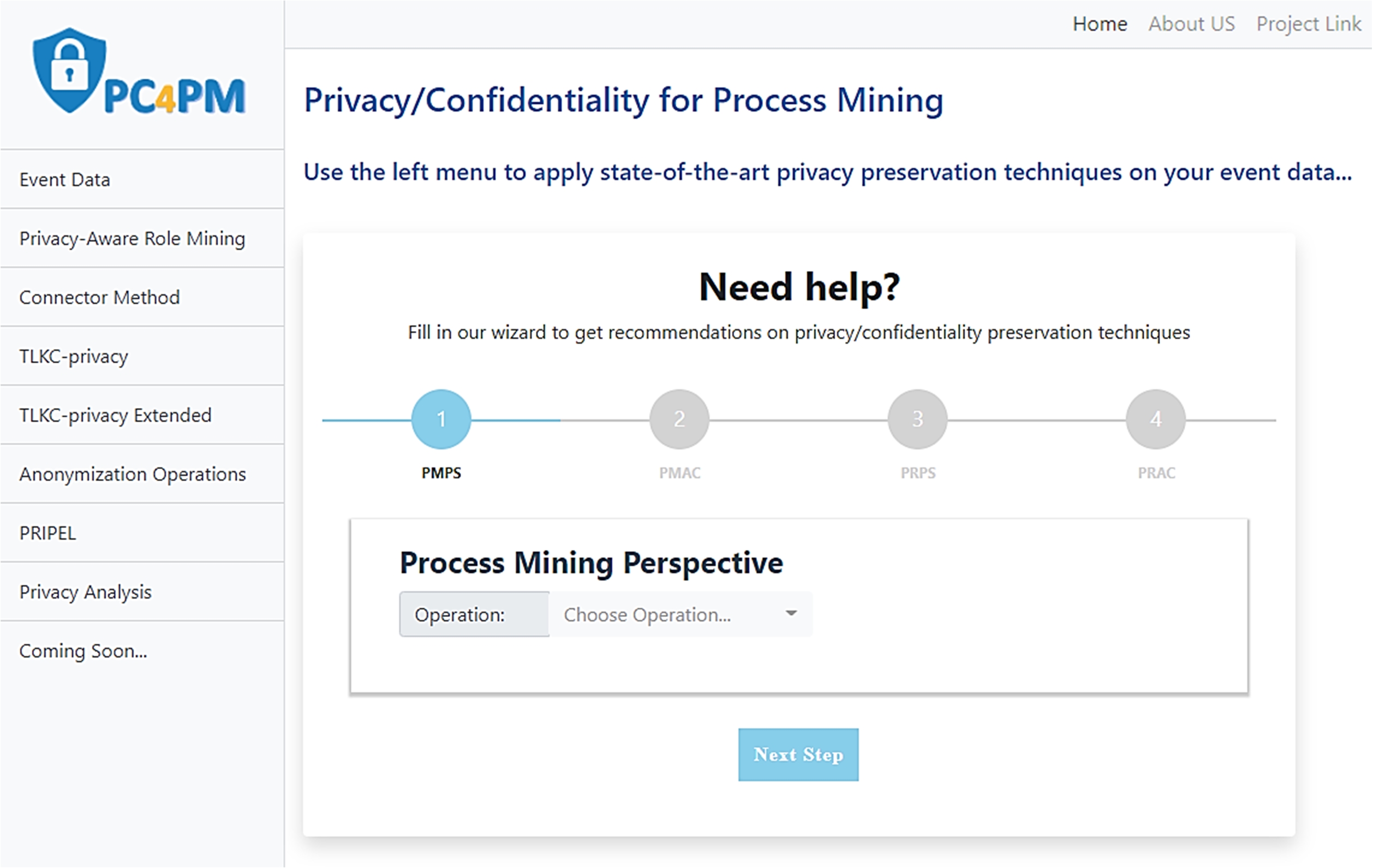}
	\caption{The home page of PC4PM.}\label{fig:home}
\end{wrapfigure}

Figure~\ref{fig:home} shows the home page of PC4PM. The left menu shows the main Django applications including \textit{event data management}, \textit{privacy-aware role mining}, \textit{connector method}, \textit{TLKC-privacy}, \textit{TLKC-privacy extended}, \textit{anonymization operations}, \textit{PRIPEL}, and \textit{privacy analysis}.
The \textit{event data management} application manages both standard XES event logs and non-standard event data, called \textit{event log abstraction} \cite{rafieippdp_google}. 
The \textit{privacy-aware role mining} application implements the decomposition method, proposed in \cite{rafiei2019role_short}, to discover roles from event logs while preserving privacy. This method perturbs the frequency of activities in an event log to eliminate frequency-based attacks. 
The \textit{connector method} is an encryption-based method for securely discovering directly follows graphs from event logs. This method breaks down traces into a collection of directly follows relations to prevent linkage attacks.

The \textit{TLKC-privacy} application implements the TLKC-privacy model providing group-based privacy guarantees for process discovery and performance analysis. 
The \textit{TLKC-privacy extended} application extends the TLKC-privacy model and considers all the main perspectives of process mining \cite{rafiei_group_arxiv}.
The \textit{anonymization operation} application, implements all the main anonymization operations proposed in \cite{rafieippdp_google} including \textit{suppression}, \textit{addition}, \textit{substitution}, \textit{condensation}, \textit{swapping}, \textit{generalization}, and \textit{cryptography}.
The \textit{PRIPEL} application presents the PRIPEL method \cite{pripel_short} which applies the notion of \textit{differential privacy} to provide privacy guarantees for event logs.
The \textit{privacy analysis} application includes three components for analyzing \textit{disclosure risks}, \textit{data utility} \cite{rafiei_quantification}, and \textit{FCB-anonymity} \cite{rafieiCEDP_arxiv}.

PC4PM supports users with a four-step user guide to help them choose the right technique(s) based on their needs. The user guidance works based on a four-dimension \textit{signature} assigned to each technique. The signature reflects the following aspects: \textit{process mining perspective} (PMPS), \textit{process mining activity} (PMAC), \textit{privacy perspective} (PRPS), and \textit{privacy activity} (PRAC).
PMPS indicates the process mining perspective that a privacy technique focuses on, e.g., control-flow. 
PMAC shows the process mining activity, e.g, process discovery, for which the utility of event data is preserved.
PRPS shows the privacy perspective of a privacy technique, i.e., resource or case.
PRAC indicates the main privacy-related activity of a privacy preservation technique, i.e., PPDP, PPPM, or PrAn. Moreover, PC4PM helps users with \textit{help tooltips} provided for the parameters used by techniques.  
PC4PM also inherits all the characteristics of its predecessor \cite{rafieippdpTool_arxiv}. Some of those are as follows: (1) Each Django application provides the results in an independent output section, (2) It enables a cycle of privacy/confidentiality preservation techniques such that the results from one technique can be added to the event data repository and used as an input for other techniques, (3) The \textit{privacy metadata} \cite{rafieippdp_google} which specify the order and type of the main anonymization operations are added to anonymized event logs.

\section{Availability and Maturity}\label{sec:available}
The source code, a screencast, a user manual, and all other resources are available in our GitHub repository: \url{https://github.com/m4jidRafiei/PC4PM}.
Each privacy/confidentiality Python package is linked to a separate GitHub project. The main GitHub project contains links to all those projects.  
In the corresponding GitHub project of each privacy/confidentiality Python package, one can find the name of the Python package, the link to the main paper, and a sample source code that shows the usage.
In terms of performance and time complexity, each privacy preservation technique which is linked to a Django application behaves differently w.r.t. the size of the input event log. Based on our experiments, the applications are able to handle real-world event logs, e.g., BPI challenge datasets: \url{https://data.4tu.nl/}. 
Moreover, all the complicated and time-consuming functions, developed in the Python packages, have a parameter to be run using multi-processing which is enabled by default. In this case, the input event log is divided into smaller pieces w.r.t. the cores of the processor hosting PC4PM.  
PC4PM is provided as a Docker container that can simply be hosted by users: \url{https://hub.docker.com/r/m4jid/pc4pm}. The Docker usage is also explained in the GitHub repository.

\section{Conclusion}\label{sec:conclusion}
In this paper, we introduced a tool for publishing event data w.r.t. privacy concerns. Our web-based tool is mainly focused on privacy-/confidentiality-preserving data publishing and privacy analysis considering both data utility and disclosure risk analyses.
PC4PM can be considered as a sanitizer that provides sanitized event logs that can be used by any process mining tool.
The architecture has been designed in such a way that other privacy preservation techniques can easily be integrated, e.g., we integrated \textit{PRIPEL} as an external library. 
The goal of PC4PM is to provide a comprehensive set of techniques that can cover all the aspects of privacy-related activities for different perspectives of process mining. We invite other researchers to integrate their solutions as independent applications into the provided framework.

\begin{acknowledgments}
  Funded under the Excellence Strategy of the Federal Government and the L{\"a}nder. We also thank the Alexander von Humboldt (AvH) Stiftung for supporting our research.
\end{acknowledgments}

\bibliography{sample-ceur}

\end{document}